\documentclass[english,pre,twocolumn,showpacs,amssymb,floatfix]{revtex4}
\usepackage[T1]{fontenc}
\usepackage[latin1]{inputenc}
\usepackage{amsmath}
\usepackage{graphicx}
\usepackage{amssymb}

\makeatletter

   \usepackage{subfigure}
 \usepackage{amsfonts}

\usepackage{babel}
\makeatother
\begin{document}

\title{Reply to the Comment on ``Passage Times for Unbiased Polymer
  Translocation through a Narrow Pore''}

\author{Debabrata Panja$^{*}$ and Gerard T. Barkema$^{\dagger,\ddagger}$}

\affiliation{$^{*}$\hbox{Institute for Theoretical Physics,
Universiteit van Amsterdam, Valckenierstraat 65, 1018 XE Amsterdam,
The Netherlands}\\ \hbox{$^{\dagger}$Institute for Theoretical
Physics, Universiteit Utrecht, Leuvenlaan 4, 3584 CE Utrecht, The
Netherlands}\\ \hbox{$^{\ddagger}$Instituut-Lorentz, Universiteit
Leiden, Niels Bohrweg 2, 2333 CA Leiden, The Netherlands}}


\maketitle

The main point raised in the Comment of Huopaniemi {\it et al.\/}
\cite{comment}
concerns the scaling of the mean time $\langle\tau_u\rangle$ it takes
a polymer of length $N$, threaded halfway in a narrow pore, to
unthread, in the absence of any external field or pulling force on the
polymer (i.e., for unbiased translocation). As argued in our paper
\cite{wolt}, the mean dwell time $\langle\tau_d\rangle$ that a
translocating polymer spends in the pore scales with polymer
length $N$ in the same way as the unthreading time. We specifically
studied this for polymers in three dimensions, whose dynamics is
described by the combination of reptation and Rouse dynamics; i.e.,
hydrodynamics is neglected.

On the theoretical side, a relevant time scale for this problem is the
Rouse time $\tau_R$, which is the longest time scale for a polymer
in bulk solution to relax in the absence of external forces. As a
function of polymer length $N$, the Rouse time scales as $\tau_R\sim
N^{1+2\nu}$. We verified that this scaling holds in the lattice polymer
used in our simulations, and also that the Rouse time is the longest time
scale for a polymer tethered to a fixed membrane \cite{longpaper}. Since
the mobility of a polymer threaded in a pore will not exceed that of an
unrestricted polymer, it follows that $\langle\tau_d\rangle\geq\tau_R$
\cite{kantor,wolt}. In the existing literature, there is {\it no
theoretical argument\/} for this inequality to reduce to an equality.
There is however {\it numerical} evidence in 2D that is inequality
is saturated in Ref. \cite{kantor}, as well as in the Comment and the
earlier works of the authors of the Comment.  With the number $s$
of the monomer located in the pore taken as a reaction coordinate, a
consequence of the above inequality is that the diffusion of this reaction
coordinate has to be anomalous, i.e.  the mean squared displacement
$\langle\Delta s^2(t)\rangle\sim t^\alpha$ with an anomalous dynamics
exponent $\alpha\leq 2/(1+2\nu)$. Again, there is no theoretical argument
why also this inequality should be saturated.

Since both $\alpha$ and the scaling $\langle\tau_d\rangle\sim\tau_R$ 
in Ref. \cite{kantor}
were obtained solely from {\it a single set of simulations to
calculate $\langle\tau_d\rangle$\/}, our first remark concerns the
{\it factual misrepresentations\/} by the Comment's authors, to
suggest that the results of Ref. \cite{kantor} are
``well-established''. The authors of the comment wished to settle
these with simulations alone.

There is plenty of numerical evidence that points towards different
scaling of $\langle\tau_d\rangle$ than $\tau_R$, both in 3D and in
2D. In our paper we reported a numerical exponent $2.40\pm0.05$ for
unbiased translocation in 3D \cite{wolt}. Another group, using a
completely different polymer model, reported an exponent $2.52\pm0.04$
\cite{dubbeldam}. In subsequent works, we have provided a {\it
full theoretical description\/} of this problem, leading to the result
$\langle\tau_d\rangle\sim N^{2+\nu}$ both in 3D
\cite{anom,longpaper} and in 2D \cite{planar}. This theoretical
description is supported by high-precision numerical simulations,
for which the 3D results we provide below in Table I.

\begin{center}
\begin{tabular}{p{2cm}|p{2cm}|p{2cm}}
\hspace{8mm}$N$ &
\hspace{8mm}$\tau_u$ &
\hspace{4mm}$\tau_u/N^{2+\nu}$ \tabularnewline
\hline
\hline 
\hspace{7mm}100 &
\hspace{6mm}65136 &
\hspace{6mm}0.434 \tabularnewline
\hline 
\hspace{7mm}150 &
\hspace{5mm}183423 &
\hspace{6mm}0.428 \tabularnewline
\hline 
\hspace{7mm}200 &
\hspace{5mm}393245 &
\hspace{6mm}0.436 \tabularnewline
\hline 
\hspace{7mm}250 &
\hspace{5mm}714619 &
\hspace{6mm}0.445 \tabularnewline
\hline 
\hspace{7mm}300 &
\hspace{4mm}1133948&
\hspace{6mm}0.440\tabularnewline
\hline 
\hspace{7mm}400 &
\hspace{4mm}2369379&
\hspace{6mm}0.437\tabularnewline
\hline 
\hspace{7mm}500 &
\hspace{4mm}4160669&
\hspace{6mm}0.431\tabularnewline
\hline
\end{tabular}
\par\end{center}

\vspace{3mm}
{\footnotesize Table I: Median unthreading time over 1,024 runs for
each value of the polymer length $N$ in 3D. Data taken from Ref.~\cite{anom}.}
\vspace{3mm}

The only numerical evidence contradicting our theory , as far as we are
aware of, is the newly produced numerical result in the Comment, and
that of Wei {\it et al.}  \cite{wei}. However, it is unclear whether
the simulations using GROMACS by the Comment's authors or those used
by Wei {\it et al.} implement purely Rouse and reptation dynamics, as
is the assumption in all theoretical work mentioned here.  Moreover,
the authors \cite{comment,wei} have taken $N$ only up to $200$, from
which an attempt to recover scaling results for $\langle\tau_d\rangle$,
in our opinion, is misleading. Specially, since one is dealing with a
numerical difference of order $10\%$, the discrepancy between different
simulation results can easily be due to finite-$N$ effects; replacing
$N$ by $N+\sqrt{N}$ or $N-\sqrt{N}$ produces double-logarithmic plots
in which the data can be fitted about equally well by straight lines,
however with exponents that deviate easily 10\% or more. We do not believe
that this apparent discrepancy can be resolved by simulations alone.

Although the full derivation of the result $\langle\tau_d\rangle\sim
N^{2+\nu}$ can be found elsewhere \cite{anom,longpaper,planar}, for
the sake of completeness we summarize it below.

Translocation takes place via the exchange of mono\-mers through the
pore. This exchange responds to $\phi(t)$, the difference in chain
tension perpendicular to the membrane; simultaneously, $\phi(t)$
adjusts to $v(t)=\dot{s}(t)$, the transport velocity of monomers
across the pore, as well! With $\Delta s(t)$ as the total number of 
monomers translocated from one side to the other in the time interval
$[0,t]$, and $\phi(t)$ playing the role of chemical potential difference 
across the pore, the two variables $\Delta s(t)$ and $\phi(t)$ are 
conjugate to each other in the thermodynamic sense. In the 
presence of memory effects, they are related to each other by
$\phi(t)=\int_{0}^{t}dt' \mu(t-t')v(t')$ via the memory kernel
$\mu(t)$, which can be thought of as the (time-dependent) `impedance'
of the system. This relation can be inverted to obtain
$v(t)=\int_{0}^{t}dt' a(t-t')\phi(t')$, where $a(t)$ can be thought of
as the `admittance'. In other words, in the Laplace transform
language, $\mu(k)=a^{-1}(k)$, where $k$ is the Laplace variable
representing inverse time. Additionally, via the
fluctuation-dissipation theorem, they are related to the respective
autocorrelation functions as
$\mu(t-t')=\langle\phi(t)\phi(t')\rangle_{v=0}$ and $a(t-t')=\langle
v(t)v(t')\rangle_{\phi=0}$.

In Ref. \cite{anom} we showed that $\mu(t)\sim
t^{-\frac{1+\nu}{1+2\nu}}\exp(-t/\tau_R)$ in 3D. This implies that the
translocation dynamics is anomalous for $t<\tau_R$, i.e.,  $\langle
\Delta s^2(t)\rangle=\int_{0}^{t}dt'(t-t')a(t')$, the mean-square
displacement of the monomers through the pore, behaves as $t^{\alpha_1}$
for some $\alpha_1<1$. Beyond the Rouse time the translocation dynamics
becomes simply diffusive. From the behaviour of $\mu(t)$ above, it is
easily shown that $\alpha_1=\frac{1+\nu}{1+2\nu}$: having ignored the
$\exp(-t/\tau_R)$ term for $t<\tau_R$, one obtains $\mu(k)\sim
k^{-\frac{\nu}{1+2\nu}}$, implying $a(k)\sim k^{\frac{\nu}{1+2\nu}}$,
i.e., $a(t)\sim t^{-\frac{1+3\nu}{1+2\nu}}$, which yields
$\alpha_1=\frac{1+\nu}{1+2\nu}$.

Thus, for $t<\tau_R$, $\langle\Delta s^2(t)\rangle\sim
t^{\frac{1+\nu}{1+2\nu}}$ and for $t\geq\tau_R$ $\langle \Delta
s^2(t)\rangle\sim t$, which together yield $\langle\tau_d\rangle\sim
N^{2+\nu}$, both in 3D and 2D. Moreover, using high-precision simulation
data, we demonstrated that in 2D the probability distribution
of the dwell time $P(\tau_d)$, behaves as $P(\tau_d)\sim {\cal
P}(\tau_d/N^{2+\nu})/N^{2+\nu}$, with a scaling function ${\cal P}(t)$
\cite{planar}.

To conclude, to date no theoretical argument has been reported for
why $\langle\tau_d\rangle$ should scale as $\tau_R$. In fact there
is a derivation why the scaling with polymer length $N$ for the
two should differ \cite{anom,longpaper,planar}. Numerical evidence
\cite{anom,longpaper,planar,dubbeldam} also points towards a scaling
of $\langle\tau_d\rangle$ different from $\tau_R$, apart from those
due to the Comment's authors and that due to Wei et al. \cite{wei}. The
theoretical formalism that yields $\langle\tau_d\rangle\sim N^{2+\nu}$
for unbiased translocation also works beautifully for translocation
mediated by a pulling force at the head of the polymer \cite{forced}, and
field-driven translocation \cite{planar,field}, providing a {\it solid
unified theoretical understanding of the dynamics of translocation},
based on the well-known laws of polymer physics.

We end our reply with the additional observation that the expression
for $\langle\tau_d\rangle$ for pore length $L$ provided in the Comment
\cite{comment,luo1}, too, is incorrect. There is general agreement
that the monomers inside the pore show anomalous diffusion with some
exponent $\alpha \leq 2/(1+2\nu)$, as discussed above. With this kind
of dynamics, the time to travel over a distance $L$ has to increase 
faster than quadratically.

\end{document}